# ERAnets

Evaluation of NETworks of Collaboration Among Participants in IST Research and their Evolution to Collaborations in the European Research Area (ERA)

## FINAL REPORT

CAROLINE S WAGNER
JONATHAN CAVE
TOM TESCH
VERNA ALLEE
ROBERT THOMSON
LOET LEYDESDORFF
MAARTEN BOTTERMAN

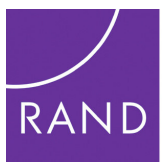 **EUROPE**

# Preface

This report presents the final results of a study, "Evaluation of Networks of Collaboration in IST Research within the European Research Area" (ERAnets), conducted for the European Commission. The ERAnets project developed and applied tools and methods to evaluate the networks of collaboration in information society technologies (IST) within the European Research Area (ERA), focusing on calls 1 and 2 of the Sixth Framework Programme (FP6). The study was conducted between February 2004 and February 2005 under the direction of Peter Johnston, Head of Unit, and Frank Cunningham, Evaluation Specialist, Evaluation and Monitoring Unit of the IST Directorate, European Commission. A steering committee of experts guided the project, including: Mark Buchanan, Author; Frank Cunningham, Evaluation Unit, DG-Information Society, European Commission; Peter Johnston, Head of Unit, Evaluation Unit DG-Information Society, European Commission; Corina Pascu, Institute for Perspective Technological Studies; May Pettigrew, Evaluation Unit, DG-Information Society, European Commission; Alberto Silvani, DG-Research, European Commission; and Bart Verspagen, Eindhoven University of Technology.

The project team included: Verna Allee, Vera Allee Associates, California; Maarten Botterman, RAND Europe, Netherlands; Jonathan Cave, RAND Europe and University of Warwick, UK; Irma Graafland, RAND Europe, Netherlands; Edwin Horlings, RAND Europe, Netherlands; Loet Leydesdorff, University of Amsterdam; Rafael Perez, University of Amsterdam; Mensur Sercovich, University of Amsterdam; Tom Tesch, Belgium; Robert Thomson, RAND Europe, Netherlands; Mark Venema, RAND Europe, Netherlands; and Caroline Wagner, RAND Europe, who was the project manager.

During the course of the study, at a workshop held in Brussels, and in other conversations, the project team received comments and suggestions that greatly aided the project. Experts who provided input include: Alex Arenas, University of Madrid; Erik Arnold, Technopolis, UK; Alberto-Laslo Barabasi, University of Notre Dame, US; Stefano Breschi, Center of Research on Innovation and Internationalization Processes, Università Commerciale Luigi Bocconi, Italy; Isabelle Collins, Technopolis, UK; Robin Cowan, University of Maastricht (MERIT); Bhaskar Dutta, Department of Economics, University of Warwick, UK; Wolfgang Glänzel, Applied Economics, Katholieke Universiteit Leuven, Belgium; Janusz Holyst, Department of Physics, University of Warsaw, Poland; Ronald Rousseau, Katholieke Universiteit Leuven, Belgium; Andrea Scharnhorst, Netherlands Institute for Scientific Information, Amsterdam, Netherlands; William Valdez, Office of Planning and Analysis, Office of Science, US Department of Energy, Washington, DC,

# Contents

# Table of Figures

# Table of Tables

# Executive Summary

Participants in the Sixth Framework Programme Information Society and Technology projects, Calls 1 and 2, when considered as project teams, create networks to share know-how and conduct research. Networks—different institutions that join together for a time-limited, specific purpose—are increasingly recognized as an important tool for knowledge sharing and innovation. They are particularly important to the European Research Area (ERA) as a way to link geographically-distant centres of excellence and to disseminate knowledge across Europe. This vision of a networked knowledge economy is central to the Lisbon Objectives.

The Sixth Framework Programme (FP6) has been structured to encourage this kind of networking across the ERA. Compared to FP5, FP6 Instruments were streamlined, incentives for collaboration were increased, and projects were increased in size. Using network analysis—tools to give insight into the effectiveness of organising for knowledge creation and exchange—this study made several important findings about the dynamics created within the ERA at the system-wide level:

- FP6 network *participants* **are more tightly interconnected** than they were in FP3, 4 and 5. In other words, there are more links in the FP6 network relative to the number of participating organisations than earlier frameworks. This connectivity offers possibilities for access to many other groups across the ERA conducting similar or complementary research.

- Other knowledge networks among private or academic groups are not well connected at the European level: When compared to these other networks, the Framework **provides an integrating function** by drawing together these participants and sectors at the European level.

- FP6 *projects* **are more closely connected** than they were in FP5; an organisation in one project is likely to be partnering participants in another project. This means that participants have a greater chance to cross-fertilise among different types of research activities with FP6.

- FP6 participants are likely to be part of other European projects such as COST or EUREKA but FP6 is far better integrated and inclusive than these other projects. This offers the chance to disseminate COST or Eureka research results, methods and perspectives to more centres.

- Connectedness of the FP6 network flows through a much larger number of alternate routes, making the network more resilient. FP6 not only makes the ERA as a whole more resilient, but increases the odds of diversified knowledge exchanges among participants, reincorporating knowledge created in other activities such as COST.

Beyond the integrating and connecting function afforded by FP6 calls 1 and 2, EC funding also provides a **complementary role** within ERA knowledge communities. A comparison of the FP6 and other ERA knowledge networks reveals a strikingly different profile, shown in the figure below (FP6 activities create the colours in the top half of the sphere and comparable knowledge networks are in the bottom half). The analysis of this data shows that FP6 networks are more likely to:

- bring together **universities and industries** into joint research projects (most non-EC funded activities are either university-only or corporate-only collaborations);

- connect **different sectors** (such as earth scientists and electronics engineers);

- include **new member states** in collaborations;

- have **more than 3 ERA countries** on a project team;

- include **patent holders** and highly-cited organisations;

- incorporate small and medium-sized enterprises **(SMEs)** in the research team.

These findings hold true across diverse research topics, from semiconductors to social inclusion, suggesting that the Framework Programme operates at a structural level rather than a topical level. Additional benefits accrue to the knowledge community rather than particular sectors.

**Figure S.1 Relative contributions of and non-FP6 networks to high-level objectives**

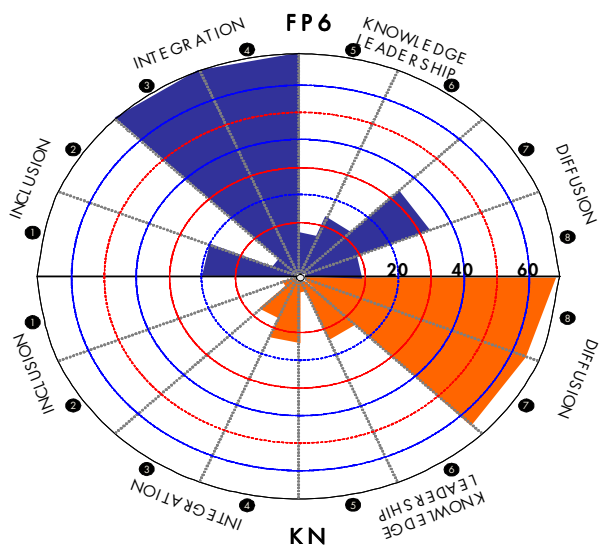

FP6 networks attract knowledge leaders from corporate and academic centres of excellence, evidenced by the rate at which relevant patent holders and highly cited institutions participate. The COST and Eureka programmes also attract corporate knowledge leaders, but the projects do not also incorporate academic leaders. Similarly, academic research networks are likely to join academic centres that are highly-cited, but they are not likely to have corporate members. The role of the European Community, at the network level, appears to be **providing an assimilation that is not otherwise operating** at the corporate, academic, or national level or within other parts of the ERA innovation system. FP6 collaborations are more likely to involve European partners and less likely to involve non-European partners, providing opportunities to re-integrate into Europe knowledge that might otherwise have flowed to other regions of the world.

Similar to studies of earlier frameworks, this study found that large organisations dominate the FP6 networks. Nevertheless, we found that the types of organisations dominating the **networks differ across the FP6 Instruments**:

- Integrated Projects are likely to have private sector organisations as central to the network;

- Networks of Excellence are centred around academic institutions;

- Specific Targeted Research Projects combine academic and corporate central members.

Central organisations in a network generally serve to organise research and to facilitate exchange of knowledge among more peripheral groups. Many organizations play a role in integrating the knowledge networks at the ERA level. While large, internally-networked organisations such as Fraunhofer and the CNRS are prominent, it is clear that many other organisations are highly influential in knitting together the participants.

SMEs participate in FP6 research networks, but they are likely to be on the edges rather than in the centre. The SMEs are connected to Instruments or projects through larger organisations. Larger organisations appear to be "gatekeepers" to Framework participation, providing stability over time; while SMEs may bring new ideas and resources they are not as likely as larger organisations to stay connected to the Framework Programme over time. A comparison of similar FP6 and non-FP6 networks (see figure) shows that SMEs are more likely to participate in Framework activities than in other types of European research networks. Their profile within FP6 is consistent with the expectation that opportunity costs associated with being tied to large research networks are relatively high, so SMEs must chose carefully when and where to be involved.

Obtaining knowledge leadership is a very significant motive for participation in FP6 projects. With 67% of participants expecting improved tools, methods or techniques, the innovations of most interest are internally focused on how people approach their work rather than externally focused on generating new products and services. This suggests that programme evaluation based only on commercial outcomes will miss the contribution to intangible asset development. The core intangible asset categories are internal structures and systems, human competence and capability and the business relationships. These intangible outcomes are perhaps where the real value lies.

Network analysis provides a systemic perspective on knowledge systems different from and complementary to survey or impact assessment evaluations. Its application is promising, and may prove highly insightful over time. Here, it shows that FP6 is more highly networked, connected integrating than earlier Frameworks, meeting at least some of the Programme's initial goals.

The Framework network may also reinforce existing relationships to some extent. This can create needed social capital and stability, but may also inhibit participation by newer members with fresh ideas. In the future, it is worth considering the possibility that network effects may also be reached with other incentives than the ones used in FP6: things like competitions or conferences. These may deserve more inquiry, since they may allow a larger number of institutions to participate at the EU level, spread the networking effects across more fields, and enable good use of links to activities outside the ERA as well as across it.

 # Introduction: New Approaches to Research Organisation

## 1.1 Evaluating FP6, Calls 1 and 2

This report evaluates research funded by the Sixth Framework Programme (FP6) Information Society and Technology (IST), Calls 1 and 2 (2002 – 2004). The evaluation was conducted between February 2004 and February 2005. The study applied network analysis to assess:

- The degree to which IST researchers in Europe collaborate with colleagues in other European countries, compared with others in their own country; and the degree to which research on the Information Society is integrated across the European Research Area (ERA).

- How the integration of IST research in the ERA changed as a result of the introduction of the new Instruments and structures for the 6th Framework Programme (2002-2006).

- Measures of the performance of co-operating alliances, as well as of the entire network across the ERA.

The EU's Sixth Framework Programme of collaborative research was designed with the expressed purpose of increasing connections across research sectors and among knowledge leaders within the European Research Area (ERA)[1] . The research is funded through five Instruments offering different organising incentives, and 22 Strategic Objectives focused on thematic research priorities. A box lists the Instruments and the 22 Strategic Objectives.

The EC's view is that the effectiveness of research depends critically on the strength of networking between research partners and across research disciplines. Policy is focused on creating tighter linkages among research units across the ERA. This process is viewed as vital to achieving the critical mass, the efficient use of resources, and for realisation of the rich web of connections that are viewed as essential to creativity.

---

[1] Council decision adopting a specific programme for research, technological development and demonstration: "Integrating and strengthening the European Research Area" (2002-2006); and Council decision adopting a specific programme for research, technological development and demonstration: "Structuring the European Research Area" (2002-2006): Available on the web at: http://www.cordis.lu/fp6/find-doc.htm

Research networks also are viewed as an effective way to meet a number of related goals, including bringing together:

- Groups from different sectors (e.g., universities, industry)
- Participants from different countries, especially smaller and new Member States
- Organisations of varying sizes, including both large and very small research and business organisations
- Researchers from different disciplines.

Although it will be years before outcomes such as patents, products, papers and participation can be assessed and even more time before we know the eventual impact of the Framework Programme on economic growth, it is possible to examine the structural patterns of networks collaboration and their processes, in order to evaluate whether the intended manner of implementation was achieved. Thus, the Instruments and other changes within FP6 can be assessed for their implementation, if not yet their goals. In this spirit, the present report describes an evaluation of research networks created by FP6 calls 1 and 2, conducted to gain insights into the impact of the new Instruments and other changes to IST research in calls 1 and 2.

## 1.2    Organisation of this Report

This introduction discusses the terms used and the methodology for the study.[2] Following this, the paper is organized into three sections presenting the context for the study, the findings, and the lessons learned:

Chapter 2.  **Context**. Following this introduction, Chapter 2 discusses the context for the study by focusing on the shifts in European Commission policy that influence the decision to move towards networked organization. The study places European policy into a broader innovation framework and discusses why networks are one approach to meeting the challenge of an increasingly competitive and globalised marketplace of ideas.

Chapter 3. **Findings**. This section presents the principle findings of the study, beginning with the system-wide characteristic of the networks created by the FP6 Instruments and Strategic Objectives. These findings are compared to earlier Framework Programmes where applicable and where data is available. The network analysis is presented as it relates to the goals of the European Commission in encouraging connections among researchers within the European Research Area. The evaluation includes indicators collected to represent the extent to which the networks created by the FP6 Instruments and the Strategic Objectives are attracting participation that can meet goals of knowledge leadership, inclusion, integration, and diffusion of knowledge.

---

[2] Other documents created during the course of this study, and published separately, provide details on the technical findings, an in-depth presentation of the methodology, a theoretical framework, and a literature review. They can be obtained from RAND Europe.

Chapter 4. **Lessons Learned**. This section discusses lessons learned from two perspectives: 1) the lessons that can be taken from the evaluation of FP6 activities for future planning; 2) the lessons learned from using network analysis as a tool for research evaluation.

## 1.3 Defining Terms

Network analysis introduces a new set of terms. It is becoming increasingly common to see references to "six degrees of separation," "small worlds," and "hubs" – all concepts that have grown out of network science. In order to understand the approach used in this study, specific meanings are needed for network concepts.

A underline{network} is a group of actors connected by some event or affiliation. In this study, networks include research organisations linked by participation in common projects. We also examined as networks research organisations connected by co-authorship; or corporations connected by a joint research venture. Within a network, actors or organisations are called underline{nodes}. In this study, the nodes in most of the networks are research organisations of some kind, such as university, industry, or publicly-sponsored research centres. Knowledge networks are groups of organisations that establish a formal collaboration to work together on a specific research goal. These can be created either out of the common interests of participating organisations in working together, or they can be created in response to incentives presented by the European Commission to link together around a specific topic. Institutional networks are groups of research institutes that operate under the umbrella of a large, coordinating unit, such as the Fraunhofer Institutes or CNRS.

Within a network, a connection between actors is called a underline{link}. A sequence of links from node to node across a network is called a underline{path}, and the shortest number of links needed to cross from one node to another is called a underline{path length}. Networks are often measured by the number of links among a specific group of nodes – if there are relatively many links (and hence, short path lengths), these groups are called underline{clusters}. Clusters thus are groups within a network where there are connection redundancies through multiple links. Redundancies can make a network highly underline{resilient} – connectivity of the cluster as a whole is not easily weakened by the random removal of links or nodes. Within networks, the relative positions of nodes can be mapped and analysed; one of these points of analysis is underline{centrality} – an assessment identifying nodes that have links to other nodes such that they are part of the shortest paths from node to node. These central nodes have an important role in holding the network together, passing information from node to node, and influencing the nature of interconnection. Indeed, a certain kind of pathway between a network is called a "small world," where even in a large network with many nodes, there can be short cuts through the network that enable information to pass quickly through what looks like a thicket of interconnections. Small worlds are emerging within network science as an important way in which knowledge is shared and information passed within large groups.

## 1.4 Methodology for this Study

### 1.4.1 Network analysis provides new insights to enrich evaluation

The request for this study was motivated in part by by breakthroughs in understanding networks. Although network analysis has been a tool for several decades, recent discoveries about the shape, structure, and evolution of networks findings emerging from the physics and biology communities have renewed interest in applying these tools to gain insight in many types of networks activities, including research. The new findings have stoked enthusiasm for using network analysis to understand highly complex systems, particularly ones that "self-organise," such as knowledge-sharing networks. The EC staff saw recent developments in network sciences as an opportunity to infuse research evaluation with new approaches and insights. Indeed, as network creation enhances the dynamism of the ERA, tools to study them must also be improved.

New tools in network analysis offer intriguing opportunities to evaluate collaborative research. Networks operate in ways that can be anticipated and their structure can be analysed using advanced techniques. When used as a tool for evaluation of collaborative research, network analysis gives insight into the efficiency of operations and knowledge dissemination. It can also show the role of different players (e.g., small and medium-sized businesses), and the openness of networks to new members. These features can influence the efficiency and effectiveness of funding and the eventual outcome of research. As EC activities create networks, the resulting dynamics can be measured and analyzed.

The best way to study networks is through social network analysis tools. Social network analysis has developed over several decades to describe the dynamics of direct and indirect interactions among purposeful and social groups. Graph theory has been used to describe the properties and dynamics of social networks that can be described by social network analysis. As noted above, these tools have been made even more interesting by recent research (described by Buchanan (2002) and Barabasi (2002)) suggesting that networks share in common mathematical and statistical features that can reveal underlying dynamics.

Within any kind of network analysis, the point of evaluation focuses on the structural characteristics of the network—this is also sometimes referred to as the typology—rather than the characteristics of the actors. This is part of what makes the analysis in this report different from other kinds of evaluation. The participants in a collaborative research project become the structural variables, or nodes, that enable the analysis of the activities as a whole, viewed as a system. The interaction and exchange within a network reveals dynamics of exchange among the participants that, in networking theory, equals more than the sum of the parts. Attempts to understand the dynamics within networks goes beyond examining inputs and outputs of research, which are the traditional tools of evaluation. While input and outputs are important they tend to be used at the beginning and end of research planning and thus they cannot help in the process of steering and managing on-going research.

### 1.4.2 Methodology for network analysis was created for this study

This study created a system-wide assessment of FP6 networks and compared them to earlier Framework Programmes as well as to comparable knowledge networks within the ERA. Analysis focused at the level of organisations and institutions within the European Research Area (rather than at the country level or at the level of the individual researcher). The scope of the study was bounded at collaborative research projects that occurred between 2000 and 2003 (depending upon the data set); as occurring within the European Research Area; and as occurring between research institutions.

During Phase I of the study, data were collected and software built and adapted to enable network analysis of the FP6 Calls 1 and 2 data. An analytic tool crafted for this project, called Nautillus, enabled analysis of networks with "institutions" or "projects" as the units of analysis. Data evaluating other IST knowledge networks were compared to the FP6 networks. These were drawn from co-authored technical publications and corporate alliances. Data on FP5 was also incorporated to allow richer analysis as a point of comparison. An example of a project-based network is shown below

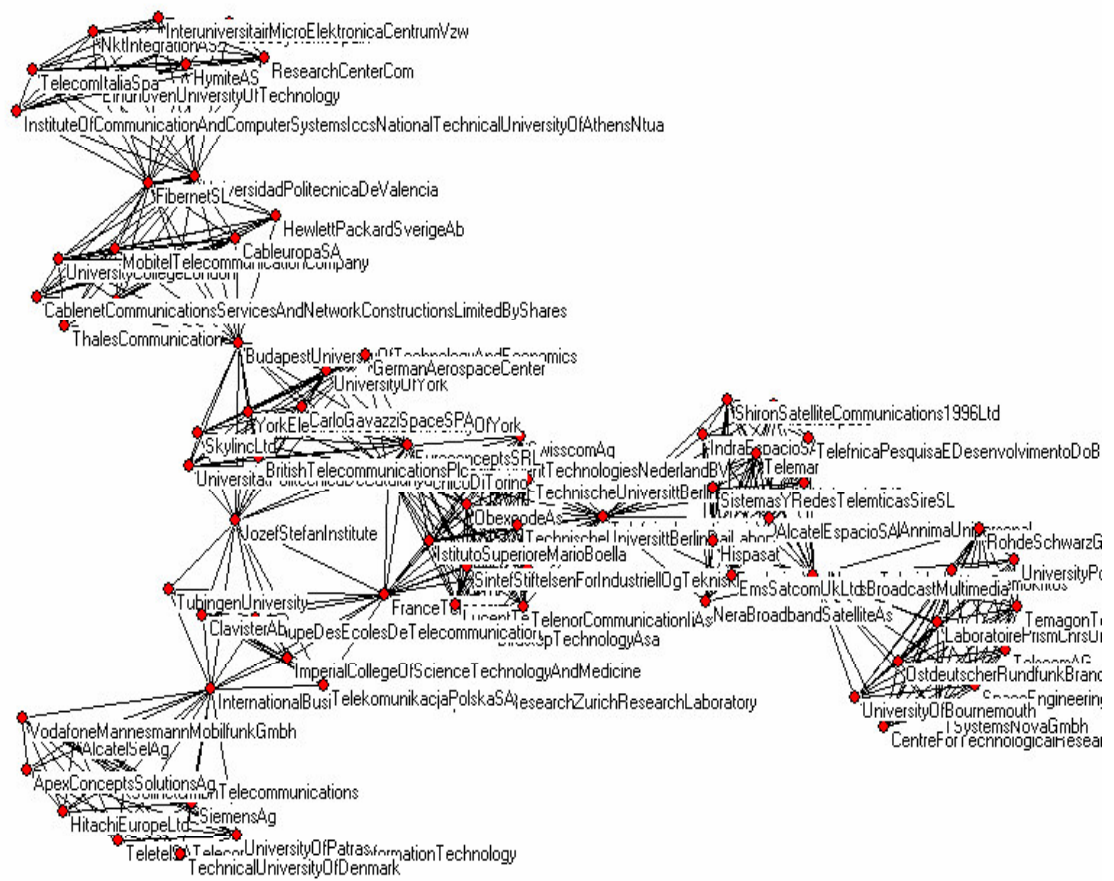

**Figure 1.1 Broadband for All – Network of Participants operating under the Specific Targeted Research Projects Instrument**

During Phase II of the study, the data were analysed and the FP6 IST RTD programme was examined. The network structures examined in this study are based on the affiliations of the participants. The affiliations are measured by participation in a collaborative research project or activity. The activities are not measured by self-reported data from surveys or other observational collection methods, but on 'quantitative' data relating to:

1. participation in EC IST call 1 and/or 2 projects;

2. participation in another European IST project (COST and Eureka);

3. co-authorship of an IST-related publication;

4. participation in a research joint venture.

Network structure and typology can provide insights into the question of whether a network tends towards integration to encourage participation of research institutes across the European Research Area. Beyond participation lie more nuanced goals of integration and inclusion, each of which was considered in the analysis to the extent possible:

- involvement of a diverse range of institutions;

- involvement of a substantial proportion of ERA institutions;

- creation of multiple linkages among institutions;

- creation of 'clusters' that can intensively exchange information and debate ideas;

- creation of a structure for efficiently disseminating knowledge;

- creating an inclusive environment to encourage participation and raise general standards and/or to correct the tendency to elitism and balkanisation in other aspects of the ERA.

To assess progress towards these goals, content analysis is required to reveal who is participating. For the purposes of understanding value and the role of incentives in influencing network structure, survey analysis is required. The network analysis moves from

- structure (typology), through

- content (quality of participation), and on to

- value analysis (intangible exchange).

At each stage, it becomes harder to provide 'hard' metrics. At the point of value analysis, interviews and surveys are needed to understand why actors participate in different Instruments and what value is realised from their participation. The project used all three types of analysis for the study: network structure, content analysis of the types of participants in the networks; and survey data to understand the motivations of those joining the networks.

**Table 1.1Strategic objectives in FP6**

| Strategic Objectives addressed in Call 1 |
|---|
| 1. Pushing the limits of CMOS, preparing for post-CMOS |
| 2. Micro and nano-systems |
| 3. Broadband for all |
| 4. Mobile and wireless systems beyond 3G |
| 5. Towards a global dependability and security framework |
| 6. Multimodal interfaces |
| 7. Semantic-based knowledge systems |
| 8. Networked audiovisual systems and home platforms |
| 9. Networked businesses and governments |
| 10. eSafety of road and air transports |
| 11. eHealth |
| 12. Technology-enhanced learning and access to cultural heritage |
| **Strategic Objectives addressed in Call 2** |
| 1. Advanced displays |
| 2. Optical, opto-electronic, & photonic functional components |
| 3. Open development platforms for software and services |
| 4. Cognitive systems |
| 5. Embedded systems |
| 6. Applications and services for the mobile user and worker |
| 7. Cross-media content for leisure and entertainment |
| 8. GRID-based Systems for solving complex problems |
| 9. Improving Risk management |
| 10. eInclusion |

 **Context: Need for new Approaches to Evaluation**

Through systematic integration over several decades (codified in the Single European Act, the Maastricht Treaty and more recently in the Lisbon Strategy) the European Union has committed to building a knowledge-based economy for Europe. The vision of Europe's leaders is an innovative economy, one that makes full use of information technologies. Collaborative research is part of the original design for this enhanced Europe, but FP6 has an even stronger focus on the integration of research in the European Research Area. As noted in the Tender document for this study: "It [FP6] has a critical role as the support for a new ERA-wide network of collaborations – which can connect most research institutes and can catalyze the more intense trans-national networking of most EU research."

To meet these goals, the Commission staff designed the FP6 Instruments to favour research teams. This didn't mean that the shape and structure of all the research teams would be the same - quite the opposite. The Instruments provide incentives for some groups to work closely together on highly innovative research, while others are expected to take on the role of exchanging information and making new connections across Europe. The 22 Strategic Objectives within calls 1 and 2 provide technical and social targets for the networked teams to focus on.

The salient features of the Instruments[3] are as follows:

- **Integrated Projects** (IPs) are very large projects with holistic workplans that connect a range of research, development and deployment activities. They have limited internal flexibility, but overall workflow is fairly well laid out from the beginning. The coordinating organisation has a key role and mediates participation – and thus has most 'bargaining power'. IPs are overwhelmingly likely to involve a wide range of organisations from the research and business communities. In some cases, work could be more modular than collaborative.

- **Networks of Excellence** (NoEs) are also large Instruments, but with much more internal flexibility to pursue 'portfolio' exploration of a range of alternatives. They are primarily intended to combine and cross-fertilise existing strands of research

---

[3] A third new instrument in the form of Article 169 projects exists where the Community may make provision, in agreement with the Member States concerned, for participation in R&D programmes undertaken by several Member States. This instrument is restricted to research initiatives beyond the scope of the Integrated Projects and the Network of Excellence and no examples currently exist in IST

around a common core issue. Their internal financing provides strong incentives for active and ongoing collaborative effort. They are more likely to involve publicly-supported research organisations and to have less centralised or hierarchical structures. NoE research is perhaps more likely to provide external (career) rewards and thus to favour 'outreach' collaboration beyond the original network.

- **Specific Targetted Research Projects** (STRPs) are suited to smaller consortia and more narrowly focused research that is innovative within a predetermined work-plan – they are more likely than NoEs to involve innovation in the strict sense but also more likely to be self-contained..

- The **Coordinated Actions** (CAs) and **Specific Support Actions** (SSAs) Instruments provide other forms of support or coordination to ongoing research efforts and areas of policy application in other Instruments.

## 2.1 Changes in Global Innovation, Knowledge Networks Affect EC Policy

The Instruments are designed to create network effects. This reflects the change in the larger research community, where developments increasingly result from connections across disciplines, sectors, and geographic distance. These changes have, in recent years, have profoundly effected innovation. Innovation and deployment rates vary markedly over time and across countries and sectors. In some cases (e.g. the recent dotcom bubble) innovative potential far outstripped actual deployment. Globalisation also affects the type of innovation – whether it is led by markets, policy or curiosity and whether it is convergent (a race for a single prize magnifies societal risk) or divergent (possibly leading to excess diversity and loss of interoperability). Finally, global networks of publication, property rights and market access profoundly affect the degree to which the benefits of innovative activity are enjoyed in the region or sector that invested to produce them. These changes alter the policy case for public research support. The overall implications are that networking matters for RTD, that policy affects network structure and therefore that structural changes affect the shape of desirable policy.

Collaboration can offer benefits of knowledge transfer and inclusion as well as contributing to innovation. Large companies can work with a number of small companies to create market opportunities that benefit both parties. Research teams that embrace new member or participants from diverse disciplines research teams appear to realize enhanced value for participants. Under-represented groups such as women-owned businesses can also be advantaged by collaborative teams. Information technology makes it possible for groups that span a large geographical area to work concurrently on a research project.

### 2.1.1 Networking as a catalyst for innovation

To understand the importance of networking, , it is necessary briefly to recap its role in the process of innovation. The development and use of knowledge is essentially a collaborative activity. It is based on:

1. the creation of new knowledge through formal or informal variation or recombination;

2.  the testing of knowledge through debate, scholarship, and market testing; and

3.  passing on knowledge through teaching, imitation, and publication.

The knowledge itself may be codified (written down), tacit (experiential) or systemic (embedded in a process), and the actors have a wide range of motives to participate, ranging from curiosity to greed. Most of the important properties of these activities are the collective; they emerge from a host of separate but linked activities.

At a more individual level, research is conducted by people. Researchers form networks based on a variety of connections differing in strength, direction, durability and purpose. For instance, co-authors of a scientific paper are linked by their shared intellectual endeavour–the paper is the evidence of a link among the authors. Researchers can also be linked - though less directly–through exchange of codified knowledge–in particular, through the peer-reviewed literature (which combines innovation, testing and dissemination in one). They are also linked through shared ideas, methodological perspectives or issues of interest. These are often shared in less formal ways, at meetings and conferences, and using informal communications like email and websites.

Innovation is increasingly tied to the ability of companies to connect across virtual links, to find new markets, to access the latest technology breakthroughs and to quickly configure working partnerships to bring products and services to market. Networks of collaboration are seen as one strategy to meet these key challenges. Analysts who study innovation point to the productivity of collective invention. The recombination of ideas in collaborative teams adds to the speed and relevance of innovation. Perhaps more importantly from a European standpoint, the international connections including Asian and American researchers can be very important sources of knowledge transfer. European researchers often are involved with foreign collaboration links, but an unresolved question is whether the knowledge is being tied back into the European innovation system. This is part of the goal of increasing network effects in the ERA.

## 2.2    FP6 Has Been Instrumental in Building Networks

FP6 support was designed to encourage the formation of networks and collaborative activities. A combination of incentives influences who participates and how they interact. The funded projects provide a sample from a wider population of actual and potential research collaborations. The EC seeks to fund pre-competitive research that provides additional and complementary support to other research already on-going within the ERA. To understand how the availability of structured support through FP6 affects networking and, through it, the productivity and other aspects of ERA, it is necessary to consider this process directly.

Project consortia are typically assembled in light of the EC work programme and calls for proposals in which the FP6 programme is more formative than responsive. While specific Instruments and Strategic Objectives are naturally more attractive to some research groups than to others, many groups will form, develop or alter their research agenda to benefit from EC financial support and networking opportunities. The selection will depend on a range of factors; the match between current research interests or competencies and those of the Programme; the opportunity costs of forming a consortium and accepting a contract; and the perceived benefits of participation.

A second element of context concerns the (unobserved) behaviour and networking structures within successful project consortia. Collaborative innovation requires both a degree of common understanding among the partners and a degree of flexibility to take advantage of emerging findings. The literature on innovation distinguishes (in lay terms) innovation that results from combination of existing knowledge and de novo innovation. Beyond this, the evolution of useful knowledge rests on the generation of new ideas, their testing and validation and their dissemination and/or application. The shape, pace and direction of research collaboration has proven remarkably sensitive to such structural elements as the institutional culture (e.g. higher education, private industry, publicly-supported research organisations, etc.), the 'distance to market' (e.g. basic, applied, application-orientated research), the motivations and property rights of the participants (e.g. future research funding, market returns, authorship rights, etc.), etc. Variations among these determine the willingness of researchers to collaborate.

Other modalities of research funding and reward (both financial and otherwise) have produced innovation networks with distinctive structural characteristics. In particular, there are fairly sharp divisions along disciplinary lines within the 'pure research' community, between public and private-sector research communities, between commercially- and policy-orientated appliers of research products and among different national and regional 'systems of innovation.' These are shifting, thanks to a range of policy actions and changing knowledge communities, but still embody a degree of fragmentation which may impair the flow and application of knowledge – particularly for problems that cut across these boundaries.

Related research coming from innovation studies has pointed out that collaborative networks are responsible for an increasing share of new developments. Indeed, it can be argued that in a knowledge-based economy, testing, diffusing and adapting ideas are at least as important as creating them, and networks are highly effective in facilitating these functions. Recognizing the strength of networks in contributing to innovation, the Sixth Framework Programme made the creation of research networks across the European Research Area (ERA) a high priority. These networks have value in themselves: they enhance the utility of other RTD support, teaching, and innovation; they further researchers' careers, and they can improve the solution of crosscutting, collective or complex problems. The hidden point is that they have different types of 'value' to different stakeholders (funders, politicians, participants, user communities, etc.) Institutions join networks to gain the benefits of combined effort. It is this part of the activity that is addressed in this report.

### 2.2.1 Networks require new tools for evaluation

The policy goals of encouraging self-organising networks to achieve goals of integration, inclusion, and innovation pose significant challenges for evaluation and monitoring. Traditional tools based on national systems of innovation, on a linear concept of research and development (from basic research to the marketplace), and on inputs (such as research spending), are inadequate to assess research within the European Research Area. Increasingly, ERA research is networked, spans disciplines and political borders, and includes participants from different sectors (such as university and industry researchers in common research projects). Each of these factors adds a measure of complexity to those seeking to do post hoc evaluation or programme planning. The challenge to the European Commission in keeping up with the significant changes in the knowledge-based economy requires new and improved tools for evaluation and assessment.

The ERA can be considered a knowledge network operating on both self-organising principles as well as resulting from specific policy decisions of the European Union. The ERA networks are embedded within other knowledge networks, specifically ones operating at local, national levels, as well as those operations at the global level. As such the ERA both draws from and contributes to knowledge networks at the local and national levels, and draws from an integrated knowledge from the global level. The dynamic and interchange can be observed and analysed using network analysis tools. Even so, the tools themselves are in the early stages of development, so this study both developed and applied the network analysis tools used to provide insight into FP6.

 # Findings: FP6 Networks Integrate Research across the ERA

The FP6 Instruments greatly increased the networking opportunities for those participating in funded projects. Compared to FP5, FP6 is characterised by a smaller number of projects that are larger in scale (see table below); this size difference enhanced interconnection among IST research institutions. A higher proportion of participating institutions were linked together through the projects funded by Calls 1 and 2. This increased the opportunity to share knowledge and resources across the European Research Area, and it may have brought together researchers who would otherwise not have had a chance to collaborate. More specifically, compared to FP5, FP6 involved fewer organisations, each of which (on average) was involved in fewer projects. Each of the projects on average involved more organisations than was the case in FP5; this suggests that they had a better chance to meet and interact with each other than if they had been in smaller projects.

FP6 Instruments influenced changes that are described in the following sections. Briefly, these changes are:

- FP6 Instruments differ among themselves but are all tightly interconnected;

- There were more interconnections among projects;

- Different sectors were linked together;

- The FP6 networks provide an integrating function within the ERA knowledge community

- The Instruments attract different but complementary participants;

- FP6 is more likely than other ERA networks to include SMEs.

**Table 3.1 Comparison of Framework Programmes**

|  | FP 3[a] | FP4[b] | FP5[c-d] | FP6[e] |
|---|---|---|---|---|
| Number of organisations in funded projects | 6291 | 5335 | 8026 | 3351 |
| Number of projects | 2131 | 1743 | 2786 | 374 |
| Average number of organisations per project | 7.1 | 7.08 | 7.16 | 15.13 |
| Average number of projects per organisation | 2.4 | 2.31 | 2.38 | 1.69 |

**a&b: Breschi, Stefano and Lucia Cusmano (2003).**

**b: FP4 data covers first part of 4th FP only (Breschi and Cusmano (2003: 9).**

**c: Johnston and Pestel (2002).**

**d: Data from European Commission (2004).**

**e: Data from European Commission (calls 1 and 2)**

## 3.1   Instruments Produce Tightly Interconnected Networks

The incentives for collaboration created by the Instruments in FP6 appear to have influenced network dynamics. The five funding Instruments differ in the priorities they attach to objectives: The most notable difference is the relatively large number of participating organisations per project under the Integrated Projects (IPs) and Networks of Excellence (NoEs) Instruments, and the comparatively small number of participants per project in the Specific Support Actions (SSAs), Coordinated Actions (CAs), and Specific Targeted Research Projects (STRPs). The Instruments that create the comparatively large networks are much more interconnected than the other networks. The SSAs and CAs are designed to support the other Instruments, so it makes sense that they will have smaller and more fragmented network structures: they are in fact increasing the strength of the other Instruments. Table 3 shows that the Instruments create networks that are very tightly connected, with short path lengths through the networks and many opportunities for connection.

There are noteworthy differences among the FP6 funding Instruments in terms of the structures of the operating networks. The NoEs are very tightly and densely networked projects, with many interconnections. The IPs have more clustering, meaning there are more groups that have formed within similar projects that in turn share participants. The STRPs are less likely to share participants across networks. All of them, however, are more tightly networked and interconnected than earlier Frameworks. While the tight connection may not be the same in each project, participants in these large networks do belong to a chain of projects each of which shares some members with others. Thus, in principle, each may be more likely to share information with others than an 'outsider' would. Because of their size, the IPs are likely to attract considerable outside interest, while the NoEs tend to involve (to some degree) a high proportion of researchers active in the 'nearest' fields.

### 3.1.1 Increased links among participants

The links among participants in the FP6 networks create connections that allow any two organisations to be only a few "handshakes" away from similar research centres working on related topics, as can be seen by the shorter path lengths shown in Table 2. This greatly increases the opportunity for researchers to connect to knowledge resources within the ERA. This connectivity goes well beyond simply gaining access to the published output of research activities. Everyone, whether an FP6 participant or not, can read the literature. However, much of the research within FP6 (and other publicly-sponsored research programmes) does not find swift expression in the open literature. Moreover, even if project results are documented in project deliverables that are available to others researchers, the necessary awareness of them may be lacking. Also, such deliverables may be written more for a policy than a scientific audience, or may require further discussion and clarification to bear fruit in others' research endeavours. In this context, indirect connections via FP6 project participation are opportunities to participate in other fora for exchange and collaborative exploration that are not available to a wider audience. In particular, these links are indicators of participation in collective events (e.g. the IST event or concertation workshops).

In this world of ideas, distances matters because people attending such events naturally gravitate towards – and participate more intensively in – events that involve people with whom they have a connection. That "distance," and therefore the ability to close the gap between any two organisations working on the FP6 activities is shorter than it has been in any of the earlier Frameworks. Of course, the shorter institutional distance does not indicate a shorter *cognitive* distance, but it shows increased possibilities for local search for new ideas.

**Table 3.2 Network characteristics of Framework Programmes**

|  | FP3&4[a] | FP5b,[c] | FP6[d] |
|---|---|---|---|
| Number of links in funded projects | 103678 | 76995 | 66242 |
| Density (x 100) | 0.2152 | 0.2391 | 1.1802 |
| Number of organisations in 1st component of network | 9455 | 7389 | 3287 |
| As percent of all organisations | 96.30% | 93.70% | 98.10% |
| Average path length* | 3.16 | 3.14 | 2.63 |
| Maximum distance/diameter* | 8 | 9 | 7 |
| Clustering coefficient* | 0.826 | 0.758 | 0.893 |

a: Breschi, Stefano and Lucia Cusmano (2003); FP4 data covers first part only (Breschi andCusmano (2003: 9).

b: Johnston and Pestel (2002).

c: Data from European Commission (2004).

d: Data from European Commission (calls 1 and 2).

*: average path lengths, maximum distances, clustering coefficients calculated for the first, largest component only.

### 3.1.2 More interconnection among projects

The FP6 network can also be examined using projects as nodes, in contrast to other analysis presented here that considered organisations as nodes. The project analysis presents a different and very interesting perspective on the relationships created within FP6. The most notable point is that, in the more technical of the Strategic Objectives, the project networks have more clustering, meaning, there is more likelihood that two projects within a Strategic Objective share a common organisation. The table below shows that the more technical a topic, the more likely it is to have organisations that participate in more then one project. We suggest that clustering is an indication of a clear subject focus within a certain strategic objective. The higher clustering coefficient may evidence convergence around a technical solution, while a low clustering coefficient may evidence varying and competing approaches. The table compares the clustering coefficients for all the Strategic Objectives. The figure illustrates the connections between projects in the CMOS Strategic Objective.

**Figure 3.1 The Clustering Coefficient of Project Networks per Strategic Objective: an indication of degree of focus?**

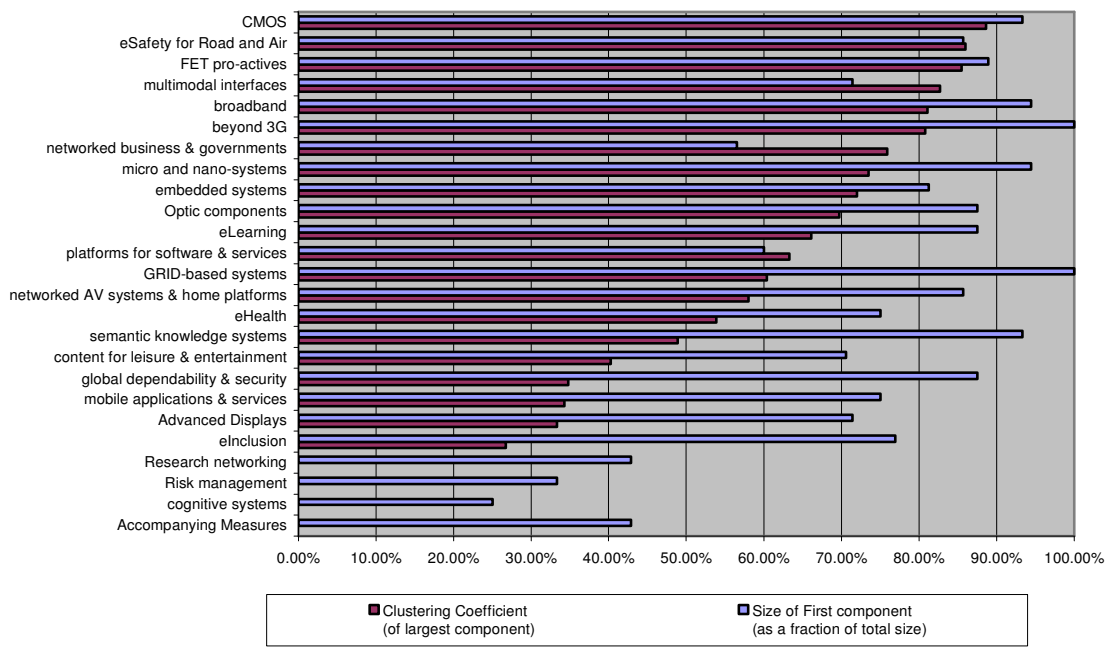

**Figure 3.2 Network of Projects Sharing Organisations within FP6 CMOS Research**

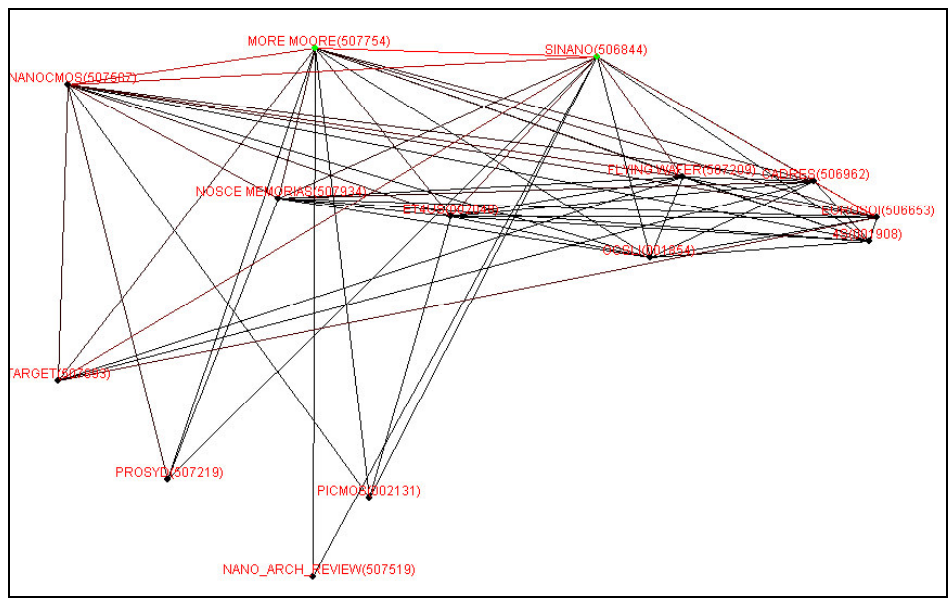

### 3.1.3 Increased connections among different sectors

FP6 played an integrating role among sectors that is not evident in other parts of the ERA. The FP6 projects were more likely to show linkages among different research sectors (e.g., universities, industries, and government research centres) than was the case in FP5. The networking opportunities offered by the FP6 Instruments increased the ability of institutions to reach across sectors and across disciplines – a key factor in innovation according to many economists. Within comparable knowledge networks not organised under the Framework, collaborations are overwhelmingly within sectors (e.g., authors of peer-reviewed journal articles are mainly academics; patents are mostly held by business.)[4]

Research collaborations within the ERA are likely to link geographically close partners either within a country or just across the border. This pattern holds for research in general, not just in the European Research Area. (As one indicator, internationally co-authored articles make up about 20 percent of all articles published. Within Europe, this percentage is higher, partially because of the influence of Europe Commission RTD support.) Given the global nature of IST research, European research centres are highly likely to link with non-European centres. Despite this, we found that the incentives provided by the Framework Programme favour intra-European research links.

### 3.1.4 More integration in Framework than in other ERA nets

Research collaborations that span the ERA operate differently from their national counterparts. National level selection processes tend to be mediated by scientific elites, typically on the basis of excellence, though social goals, including regional and future

---

[4] Cross-sectoral collaborations may publish results under the aegis of the academic partner, and patent under the aegis of the firm, based on the idea that the Intellectual Property should go to the entity best able to exploit it. This kind of distinction is difficult to make with the data examined in this study

benefits play an increasing role. The networks at the national level are both more fragmented (in part through disciplinary specialisation) and less trans-European in character. In contrast to networks supported primarily by national funding, the ERA-wide networks supported by FP6 are organised around influential semi-public organisations (groups such as Fraunhofer or TNO, which themselves are networked) that act as an 'attractors' for smaller and more peripheral groups. The network is further shaped by European Commission selection processes that combine technical excellence and social goals (with relatively more weight on the latter) as criteria for participation.

We compared the structures of FP6 networks to those of other European knowledge networks to look for the added value of the Framework. This analysis showed that FP6 is reducing the fragmentation of research communities at the ERA level. More specifically, there is more intra-European collaboration and less collaboration with non-European organisations. Clearly, FP6 provides a vital factor pushing existing national knowledge networks towards European integration. This works on two levels: organisational bridging is fostered by the funding Instruments and a focus on transnational and transdisciplinary issues is provided by the strategic objectives. In addition, by fostering intra-European collaboration, FP6 provides the opportunity to re-integrate into Europe knowledge that might otherwise have flowed to other regions of the world. It is worth considering whether European integration inhibits the realisation of economic gains by creating innovation barriers at the borders of the EU. Such a finding does not emerge from this study, but it should be considered for future research.

**Table 3.3 Comparison of FP6 Instruments**

|  | Integrated Projects | Networks of Excellence | Specific Support Actions | Coordinated Actions | Specific Targeted Research Projects |
|---|---|---|---|---|---|
| Organisations in funded projects | 1575 | 928 | 250 | 236 | 1129 |
| Number of SMEs | 372 | 99 | 77 | 45 | 320 |
| Number of projects | 97 | 42 | 44 | 21 | 170 |
| Number of edges/links | 31647 | 26169 | 1504 | 2045 | 6429 |
| Density (x 100) | 2.5532 | 6.084 | 4.8321 | 7.3747 | 1.0096 |
| Components in graph | 1 | 1 | 28 | 10 | 18 |
| Organisations in 1st component | 1575 | 928 | 71 | 150 | 1071 |
| As percent of all organisations | 100% | 100% | 28% | 64% | 95% |
| Average path length* | 2.47 | 2.11 | 2.23 | 2.59 | 3.64 |
| If random (clique) | 2.45 | 2.05 | 3.08 | 2.53 | 4.04 |
| Maximum distance/diameter* | 4 | 3 | 4 | 5 | 9 |
| Clustering coefficient* | 0.908 | 0.899 | 0.96 | 0.959 | 0.91 |
| If random (clique) | *0.013* | *0.030* | *0.024* | *0.037* | *0.005* |

Compared to other EU projects such as COST, FP6 also offers more networking opportunities that tie research back into the European level. COST data examined for this study covered 498 organisations in 29 projects. However, only a few organisations are involved in more than one project, limiting potential networking benefits from collaboration. When the FP6 participants are added into the mix, the COST programme, and by extension other programmes, are tied back into a larger knowledge community.

The COST network has what are called 'cut-point' organisations. Their participation ties the network together (weakly), but this connectivity would disappear if they were removed. By contrast, FP6 networks are connected through many more alternate routes, making the network more resilient. FP6 not only makes the ERA as a whole more resilient, but increases the odds of diversified knowledge exchanges among participants, reincorporating knowledge created in other activities such as COST.

When FP6 networks are 'added' to other ERA knowledge networks, they do not appear to duplicate existing patterns of interaction and thus extend and connect the ERA as a whole. Of course, many networks show skewed distributions of connectedness, with a central core of highly linked organisations and a periphery of less well-connected entities. FP6 networks are perhaps less skewed than patent, publication or alliance networks. The highly-central nodes in project-based networks (e.g. FP6 and COST) are often large institutions from 'peripheral' Member States, whose inclusion reflects a preference for national diversity, the relatively concentrated relevant research sector in smaller member states and the (consequent) concentration of specific knowledge and working practices in a handful of institutions[5] . Core organisations in patent networks are likely to be large industrial concerns best placed to exploit IPR in the marketplace and thus likely to wield more direct power. The 'central' players in publication networks are likely to be University centres of excellence in a given field or large research centres housed at such Universities. It is interesting that the ERA organisations with the greatest centrality according to the Strategic Objectives in the case studies are not necessarily those dominating broader research and education 'league tables.'

One insight gained from viewing the Instruments as networks is this: they are not dominated by just a few large organisations. Many organizations play a role in integrating the knowledge networks at the ERA level. While the large internally networked organisations such as Fraunhofer and the CNRS are prominent in these networks, it is clear that many other organisations are highly influential in knitting together the participants. Analysis shows that the type of organisation central to the network -- and therefore "influential" in structural terms -- differs among the various Instruments. The Networks of Excellence have a group of academic institutions that play a central role, including Eindhoven University, University of Madrid, the French CNRS system, the Technical University of Denmark, and so on. The central players in Integrated Projects are more likely to be industrial and research institutes such as Alcatel, Thales Communications, Philips, British Telecom, and so on. The figures below provide a network illustration where the reader can examine the different organisations that are central to the IPs and the NoEs.

---

[5] This specialised knowledge is both substantive and procedural – not all institutions can meet stringent accounting and managerial requirements for EC funding and there are advantages in specialisation if local subcontracting arrangements permit

**Figure 3.3 : Institutions Participating in IPs with a Betweenness of >1**

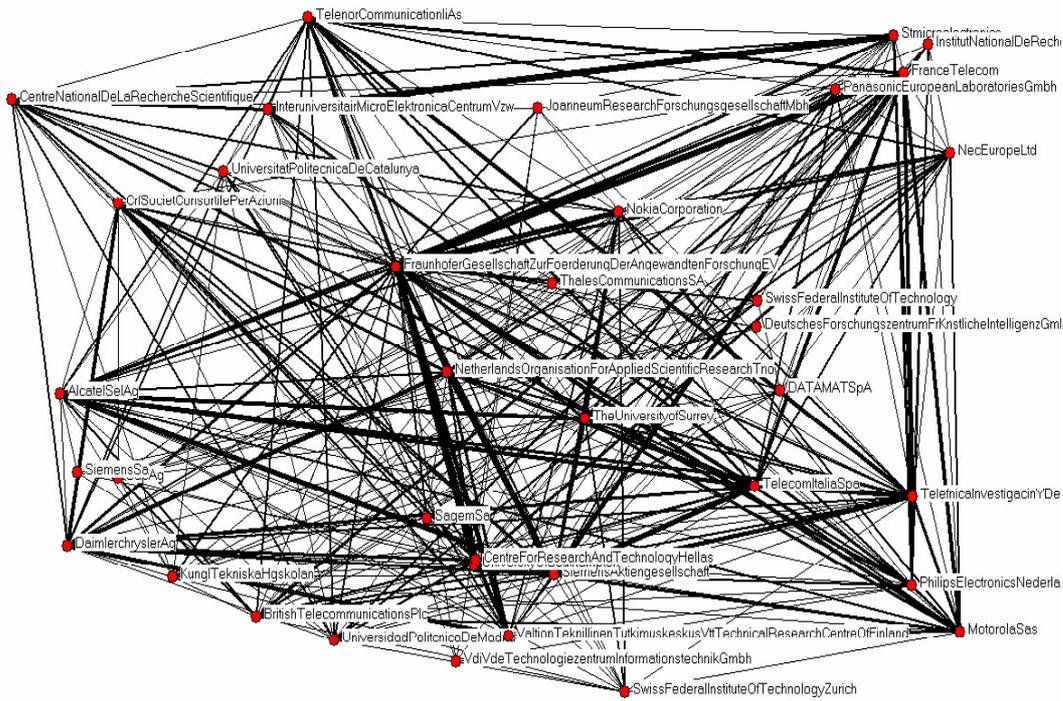

**Figure 3.4 Institutions Participating in NoEs with a Betweenness of >1**

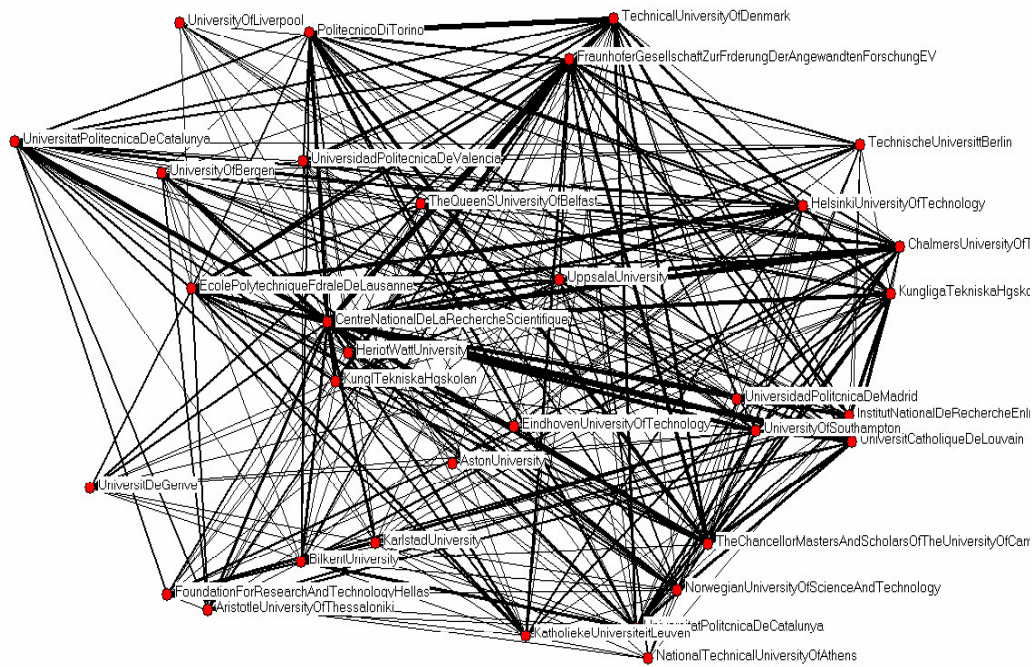

### 3.1.5  SMEs participate through larger organisations

Network analysis also provides a view into the role of small and medium-sized businesses (SMEs) in the Framework.  As might be expected, large organisations and SMEs play significantly different roles.  The SMEs are not central participants in the networks; as a rule, they do not act as hubs drawing other organisations together.  They generally enter the networks in collaboration with larger organisations.  Their profile within FP6 is consistent with the expectation that, for smaller businesses, the costs associated with being tied to large research networks is relatively high, and therefore SMEs must chose carefully when and where to be involved.  Working with larger organisations helps to bring down some of the transaction costs of writing proposals and maintaining communications with other institutions.

The population responding to the first call in many ways resembled that from previous FPs – rich in SMEs and in numbers of proposals. As a result of the shift to fewer and larger projects, many of these were unsuccessful (especially the SMEs) and the first call retained projects reflect the conscious choices both of those forming consortia and of the selectors. The second Call was influenced by these selection processes in various directions. Perceived odds of retention probably dropped, which would have discouraged some proposals and research organisations and sharpened the focus and selectivity of others as regards both participants and choice of Instruments. In particular, the climate would have become more difficult for organisations (particularly SME research organisations) that had become accustomed to a steady flow of work in small FP projects . This reflects three factors: increased prevalence of co-funding (such organisations are rarely able to use the additional cost model); asymmetry of negotiating power (the new larger Instruments depend critically on a core of strong, managerially competent 'hubs') and decreased odds of success, which would raise the variance of income flows. For some, this would discourage further participation; for others the interruption in income would have made second-round proposals even more crucial and increased the leverage of strategic objectives and instrument design. The odds of success would likely have interacted with the increased prevalence of co-funding in a similar way: diminished odds of follow-on or related work would cause some to turn away from the Framework Programme, while others would feel a stronger incentive to obtain co-funding. The data do not permit a clear analysis or modelling of these combined effects, but they should be borne in mind when interpreting the changes in structure associated with the evolution of FP6.

## 3.2  The Strategic Objectives, Calls 1 and 2, have a Structural Impact

The FP6 IST Strategic Objectives within Calls 1 and 2 focus on content or community goals of interest to the European Commission.  The Strategic Objectives and associated calls for proposals encouraged institutions to self-organise into teams for the purpose of specific, targeted research.  For FP 6 Calls 1 and 2, 22 Strategic Objectives were delineated, with each one offering a range of different Instruments as an organizing function to address the goal.  Thus, the selection of an organisational form was guided by the Call wording and the associated budget allocation.  The policy decisions are likely to have had some impact in shifting the 'natural' or default way in which participants would organize their networks–even though participation is left to the team itself.   Self-selected teams make proposals to the European Commission; which funds a subset of the proposals. The resulting projects (and the roster of participating institutions) create the basis for network analysis.

The retained projects within the FP6 Strategic Objectives create densely connected and tightly linked networks. Compared to the non-FP6 networks, they are much more integrated across Europe, tightly connected and cross-sectoral. These contrasts in structure and emphasis suggest that FP6 is serving an integrative function that is not supplied by other ERA knowledge networks. The first part of this chapter compared FP6 to other European knowledge networks to draw out details of the roles of different communities in ERA IST research. In addition, as will be discussed below, the Strategic Objectives create networks that have "small world" properties: analysts have suggested that these types of networks are particularly good at sharing and diffusing knowledge within a specific community. Finally, it is clear from survey data that FP6 participants expect significant value from participating in the Programme. This is discussed in this section as well.

At the level of the Strategic Objectives, the structure of the networks is not as revealing as network analysis at the system-wide level, in part because the networks are much smaller. Thus, for the Strategic Objectives, the focus is on the quality of participation. To better evaluate networks arising within FP6 Strategic Objectives and understand the quality of information available, six Strategic Objectives were examined in detail and compared with networks among IST researchers within the ERA that are not organised around the Framework Programme. The six cases were chosen to represent three themes that broadly characterize the topics represented by the Strategic Objectives in calls 1 and 2. These are shown below.

**Table 3.4 Case study strategic objectives**

| Technology |
| --- |
| Broadband for all |
| Pushing the limits of CMOS, preparing for post-CMOS |
| Security and Assurance |
| Towards a global dependability and security framework |
| Improving risk management |
| Social and Community Issues |
| eInclusion |
| eHealth |

To view the extent to which the FP6 networks are complementing other European activities, comparable knowledge networks were constructed using data from journal literature, patent records, and databases of joint research ventures within Europe. The data were sorted into relevant subjects using keywords derived from the FP6 background materials. It is important to note that these comparable knowledge networks do not exclude FP6 participants – in fact, there are extensive overlaps. In part, these data provide different views of the underlying ERA network, as well as indicators of other forms of collaborative activity. The comparison between the different communities gives an indication of the extent to which FP6 networks integrate activities across the ERA.

For each case, data drawn from Calls 1 and 2 were used to generate FP6 networks. To put the FP6 activities into perspective, data sets on other types of collaboration were combined to create comparable European IST knowledge networks. (We call these "comparable knowledge networks.") In order to acquire comparable data on other knowledge communities with the most relevant experience, materials were chosen based upon key-words drawn from the respective FP6 thematic area descriptions. To provide additional insights into the dynamics of these two networks, data on eight indicators were also calculated and analysed. These comparisons shed light on the participation of industrial and academic knowledge leaders. Knowledge leadership was determined was determined by directly relevant patents and/or directly related cited articles. We also compared the extent of cross-sectoral (university-industry) links and participants from new member states. FP6 and non-FP6 communities were compared using the indicators shown below.

**Table 3.5 Data collected on both FP6 and non-FP Research Commninities**

| | Data collected on both FP6 and non-FP Research |
|---|---|
| 1 | Percentage of collaborating partners from new |
| 2 | Percentage of projects/collaborati that have at least 3 ER countrie listed in the address |
| 3 | Percentag of projects/collaborati that have participan fro both university and industry research |
| 4 | Percentage of ERA collaborating partners holding |
| 5 | Percentag of collaborati partner with cited article in a directl related |
| 6 | Percentage of project with universit collaborato (fro any country not limited to |
| 7 | Percentage of projects/collaborations with non- |
| 8 | Percentage of SMEs participating in |

Overall, the findings show the following outcomes, each of which is described in more detail in sections below:

- The FP6 and comparable knowledge networks are complementary and have different emphases

- Both communities attract knowledge leaders

- FP6 networks are more likely to have cross-sectoral partnerships

- FP6 networks are more likely than comparable networks to include new member states

- The FP6 networks appear more likely than other knowledge networks to include the small and medium-sized businesses

- Some of the FP6 networks have "small world" properties as that term is understood in network analysis

- The FP6 participants expect to receive significant value from participating in the Programme.

### 3.2.1 Complementary knowledge communities emerge

These non-FP6 networks show considerably different characteristics. Specific types of network (e.g., collaborations within the ERA evidenced in peer-reviewed journal literature) are much more fragmented than the FP6 networks. In the following section, we discuss six case studies comparing the FP6 networks to a composite comparator based on publications, patents and research joint venture data. These networks are much less well-connected. Each cluster of linked research groups tends to be quite small compared to the institutional networks operating in FP6. They do not offer the additional integrating function across the ERA provided by membership of the FP6 'community of interest' and the events, exchanges and discussions taking place around its ongoing activities. Rather, integration is provided by ERA self-organisation and affiliation is perhaps more asymmetric and discretionary.

The comparisons reveal strikingly different profiles. The figure below shows the results derived from averaging the data from indicators across the six case studies. Both the FP6 and non-FP knowledge communities are represented. The figure provides a visual representation of the data, showing the different and largely complementary performance of the two communities across the four large goals of inclusion, integration, knowledge leadership, and knowledge diffusion. FP6 networks are stronger in integrating across the ERA and encouraging inclusion. The FP6 network exceeds the non-FP network in attracting knowledge leaders from industry. The non-FP6 networks are stronger at knowledge diffusion.

**Figure 3.5 Relative contributions of and non-FP6 networks to high-level objectives**

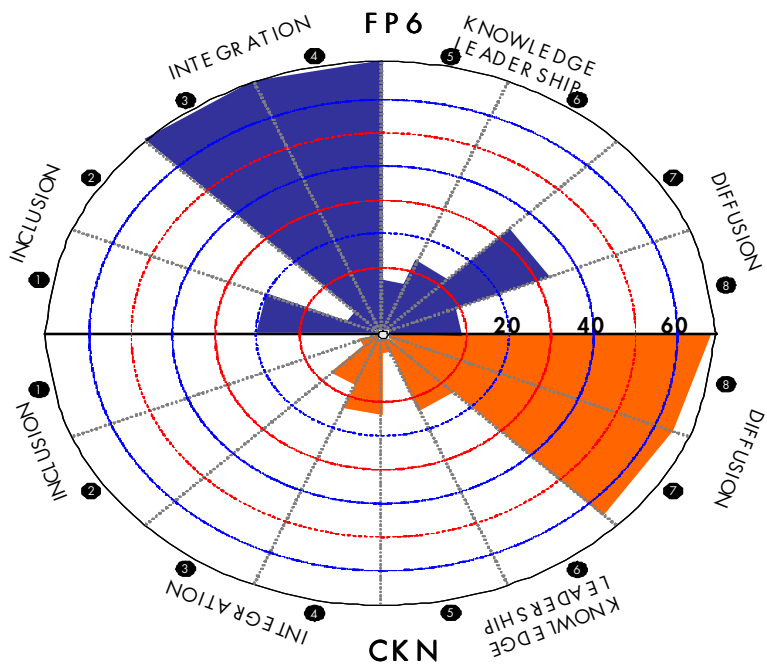

### 3.2.2 Both "communities" attract knowledge leaders

The analysis shows that the FP6 networks attract participation of prominent knowledge holders as represented by patent holders and cited institutions. The similarities end there. The FP6 networks far outstripped the non-FP6 networks in linking universities and industry, bringing participants from different ERA countries together and involving new member states. Note that, although data on the non-FP6 networks do not indicate the participation of small and medium-sized businesses, partial data and other literature indicate that the proportion of SMEs in these collaborations is much smaller than in the FP6 networks.

The table shows that for 5 of the 6 Strategic Objectives, both the FP6 and the non-FP6 networks attract the participation of knowledge leaders. This can be seen by the proportions of patent holders and organisations with cited articles. Risk Management is the exception, with no ERA patent holders participating in the FP6 research projects. Both FP6 and non-FP6 knowledge communities attracted cited institutions. FP6 eInclusion is the exception in that case. Note that some of this is directly due to the way the ERA responded to the Instruments - industry university links are almost a given for Integrated Projects and much less likely in Network of Excellence.

**Table 3.6 Case study comparisons between FP6 and non-FP6 networks**

| Case | Number of organisations in set | Relevant Knowledge Community | ERA patent holders participating in the data set | ERA institutions with cited articles participating in the data set | University industry links as a share of all projects in the data set | International collaborators within ERA as a share of total projects | Collaborators who are non EU | Collaborators who are universities | Collaborators who are from new member states | Participants who are small-med businesses (SME) |
|---|---|---|---|---|---|---|---|---|---|---|
| CMOS | 210 | FP6 | 4,2% | 31,4% | 67,0% | 93,3% | 9,5% | 48,6% | 3,3% | 10,5% |
|  | 1668 | non-FP6[a] | 3,9% | 29,0% | 28,1% | 6,8% | 66,7% | 55,4% | 0,8% | >1%[b] |
| Broadband | 272 | FP6 | 13,0% | 21,5% | 94,0% | 100,0% | 14,0% | 36,8% | 4,0% | 15,4% |
|  | 836 | non-FP6[a] | 8,4% | 26,7% | 12,6% | 8,1% | 66,0% | 63,8% | 1,3% | >1%[b] |
| GDpSec | 276 | FP6 | 14,1% | 6,7% | 94,0% | 100,0% | 9,4% | 33,0% | 6,2% | 20,7% |
|  | 3326 | non-FP6[a] | 9,4% | 34,9% | 10,2% | 8,5% | 59,7% | 66,4% | 1,0% | >1%[b] |
| Risk Management | 93 | FP6 | 0,0% | 1,5% | 78,0% | 100,0% | 8,6% | 18,3% | 3,2% | 25,8% |
|  | 751 | non-FP6[a] | 0,0% | 26,2% | 9,3% | 8,6% | 60,8% | 60,2% | 0,8% | >1%[b] |
| eHealth | 222 | FP6 | 7,0% | 12,9% | 70,0% | 100,0% | 8,1% | 27,5% | 6,8% | 23,9% |
|  | 242 | non-FP6[a] | 0,7% | 43,5% | 5,1% | 12,8% | 69,3% | 69,0% | 2,1% | >1%[b] |
| eInclusion | 172 | FP6 | 11,1% | 0,0% | 69,0% | 100,0% | 9,3% | 34,9% | 4,1% | 18,6% |
|  | 86 | non-FP6[a] | 0,0% | 95,0% | 12,9% | 3,2% | 83,5% | 74,4% | 0,0% | >1%[b] |

[a] non-FP6 = comparable knowledge (publication) network;

[b] Proportion of SMEs in the non-FP6 networks not directly measured but judged to be less than 1% of participants

### 3.2.3 FP6 networks have more cross-sectoral partnerships

The FP6 networks are much more likely than the non-FP6 networks to include partners from both universities and industry. This suggests that FP6 provides a valuable complement in meeting goals of knowledge diffusion and cross-sectoral integration. The FP6 networks are also much more highly integrated across Europe, being more likely to include participants from 3 or more countries.

### 3.2.4   FP6 networks more likely to include new member states

They are also much more likely to include participants from new member states and in this sense at least are more likely to be meeting goals of inclusion than non-FP6 networks. Despite the difficulties surrounding their participation, it appears that SMEs are much more likely to participate in FP6 networks than in comparable non-FP6 networks.

### 3.2.5   FP6 networks appear more likely to have SMEs as members

Within the data set of organizations participating in FP6, 26 percent (855) list themselves as small and medium enterprises (SMEs)[6]. An additional 181 organisations (not included in the 855 referred to above) report themselves as SMEs in some project proposals and as non-SMEs in others. In the non-FP networks, it was very difficult to determine which of the participating organisations were SMEs, since, unlike the EC data, the journal literature and corporate joint venture data do not indicate if a firm is an SME.

In examining the FP6 networks, it becomes evident that SMEs may have a difficult time breaking into the network. The extent to which the dominance of established relationships influences links also means that peripheral or new members may be locked out of the process. If existing relationships influence the ability of new members to join a network—a dynamic that appears to be at work here—then the more central and linked an organisation becomes, the more power they have to control who enters the network. If the large organisation is already saturated with relationships with existing groups, this may mean that it would be considerably difficult for new, smaller, or peripheral members to join the network. The fact that small businesses are a decreasing part of the network may be related to this phenomenon.

Despite this drop-off in SME participation in FP6, it is apparent that the Framework is much better than other research networks at including SMEs. Although data is not directly available from the non-FP networks as to the percent of participants that are SMEs, based on our judgment from examining the databases of participants, the number of SMEs participating in the non-FP networks is very small, and much smaller than the 15-20 percent participating in the FP networks.

### 3.2.6   Non-FP6 networks have non-European links

The non-FP6 networks are more likely than the FP6 networks to involve links with non-EU collaborators. This may be a boon to the ERA participants in that research, since they may be gaining access to world-class research. However, to the extent that the collaboration only involves one European team member who is not otherwise connected within Europe, diffusion of knowledge within Europe cannot be assured by these collaborations. In fact, knowledge may be flowing out rather than into or within Europe as a result of these collaborations. The FP6 activities can be seen as a way to reintegrate international knowledge back into the ERA, making it more available at the local level within Europe.

---

[6] It should be noted that there is some inconsistency in the registration of SME status that could not be fully normalised in the data sets

### 3.2.7 Framework provides an assimilation function

The patterns noted in the preceding sections persist across diverse subjects such as technology, security, and social interest, suggesting that the Framework Programme has consistent structural impacts. These findings hold true across diverse research topics, from semiconductors to social inclusion. This suggests that the overall function being played by the Framework Programme operates at a structural level rather than a topical level. The additional benefits are at the level of the knowledge community rather than providing targeted aid to particular sectors (even though these sectors may benefit).

FP6 appears to create opportunities for integration and inclusion, as well as offering some opportunities for local search, knowledge diffusion and knowledge exchange across all the IST areas (through participation in dense and highly-connected Framework communities of interest) that are not otherwise a feature of non-FP6 networks. This is even true of thematic areas related to technology, where it would be reasonable to expect the FP6 and non-FP6 networks to have some similar features, they are quite different.

## 3.3 Some FP6 networks have 'small world' properties

Many of the non-FP6 'research communities' tend to be somewhat sparse and diffuse. (This is discussed in more detail below.) By contrast, the FP6 networks show a relatively high degree of small world clustering. A small world network is characterised by short path lengths (meaning that participants are likely to be closely linked) and high clustering (meaning that participants with a neighbour in common are likely to be involved in a common project). Knowledge transmission within a small world cluster is likely to be very efficient, and the rich pattern of interconnections means that ideas are likely to be examined from a range of viewpoints. On the other hand, percolation of ideas through a network with many small world clusters may be somewhat slow.

The following tables compare various networks to a benchmark random network[7] with the same number of nodes and links. A small world network has about the same (or smaller) average path length and much higher clustering. The table below shows the degree to which the 'main clusters' (the largest connected group of organisations) associated with the various strategic objectives meet the small world condition.

---

[7] The benchmark expected values of path length and clustering are based on a network formed by random addition of new links among a fixed set of nodes.

**Table 3.7 Small worlds among FP6 networks by Strategic Objective**

| Strategic Objective | Path length | Length if random | Clustering factor | Clustering if random | Small world? |
|---|---|---|---|---|---|
| 1.1 CMOS | 1,9416 | 1,8394 | 95,44% | 8,73% | Yes |
| 1.10 eSafetyRoadAir | 1,9217 | 1,6993 | 93,14% | 11,07% | No |
| 1.11 eHealth | 2,5662 | 2,3026 | 94,43% | 5,68% | No |
| 1.12 TEnLearnCulture | 2,1743 | 1,7959 | 96,37% | 7,39% | No |
| 1.2 MicroNanoSys | 2,2645 | 2,057 | 95,81% | 6,10% | No |
| 1.3 Broadband | 2,1268 | 1,8888 | 92,46% | 7,30% | No |
| 1.4 MobileBeyond3G | 1,936 | 1,8232 | 88,29% | 7,83% | Yes |
| 1.5 GlobalDepSec | 3,2094 | 2,1673 | 96,05% | 4,96% | No |
| 1.6 MultiModal | 1,958 | 1,8228 | 94,29% | 10,15% | Yes |
| 1.7 SemanticBsdKnow | 2,5291 | 2,156 | 93,63% | 6,41% | No |
| 1.8 NetAudioHome | 2,1086 | 1,9956 | 91,67% | 7,01% | Yes |
| 1.9 NetBusGov | 1,9312 | 1,6771 | 96,89% | 11,37% | No |
| 2.1 AdvDisplays | 2,0804 | 2,0228 | 95,80% | 13,31% | Yes |
| 2.10 eInclusion | 2,3497 | 1,9339 | 96,58% | 9,04% | No |
| 2.2 OpticalPhotoComp | 2,153 | 2,0078 | 94,66% | 7,26% | Yes |
| 2.3 OpenDevSofSer | 2,1612 | 2,1756 | 93,64% | 8,17% | Yes |
| 2.4 CognitiveSystems | 1,4396 | 2,0414 | 96,34% | 26,02% | Yes |
| 2.5 Embedded Systems | 2,1568 | 2,0337 | 92,15% | 7,51% | Yes |
| 2.6 AppSerMobile | 2,6371 | 2,175 | 96,53% | 6,09% | No |
| 2.7 CrossMedia | 2,6965 | 2,4956 | 95,65% | 5,51% | Yes |
| 2.8 GRID | 2,1579 | 1,9383 | 94,08% | 9,34% | No |
| 2.9 ImprovingRiskMgM | 1,6939 | 1,721 | 97,10% | 20,30% | Yes |

Two features are immediately evident. First, the Strategic Objectives that fail to meet the small world criterion always do so because of path length: organisations sharing common neighbours are almost invariably co-participants in at least one project. Second, the second-Call Strategic Objectives have a much greater prevalence of small worlds, possibly indicating a tighter consortium structure in response to the unexpected selectivity of the first Call.

**Table 3.8 Small Worlds among FP6 Networks by Instrument**

| Instrument | Cluster | Path length | Length if random | Clustering factor | Clustering if random | Small world? |
|---|---|---|---|---|---|---|
| CA | 1 | 2.9335 | 2.1407 | 97.05% | 6.81% | No |
| IP | 1 | 2.6094 | 2.4865 | 90.86% | 1.20% | Yes |
| NoE | 1 | 2.1861 | 2.0905 | 90.34% | 2.74% | Yes |
| SSA | 1 | 1.8798 | 1.6846 | 97.27% | 18.81% | No |
| SSA | 2 | 1.9909 | 2.0659 | 96.50% | 12.31% | Yes |
| SSA | 3 | 1.4615 | 2.1066 | 96.15% | 25.00% | Yes |
| STRP | 1 | 4.0314 | 3.9721 | 90.90% | 0.55% | Yes |
| STRP | 2 | 1.4737 | 1.8614 | 97.37% | 25.00% | Yes |

The small world impression is reinforced when we consider the networks by Instrument. All Instruments show very high clustering. Because the networks are quite large when aggregated across Strategic Objectives, the expected clustering in the random network comparator tends to be relatively small. On the other hand, path lengths are fairly close to the benchmark expectation in most cases: the only reason the Coordinated Actions and the main cluster of the Specific Support Actions fail to be considered small worlds is because their path lengths are relatively long, representing a dispersed 'backbone' of key organisations.

Many analogous non-FP6 networks are also small worlds according to these criteria. The main exceptions are the COST network and patent networks for the GlobalDepSec and eInclusion strategic objectives. The latter are anomalous in that they do not have many more links than nodes – they are more star-like and less clustered than the FP6 counterparts. In the case of COST, the key structural difference is a long chain of dense clusters sharing a very few key organisations who nonetheless differ from cluster to cluster. This leads to a higher average path length – in other words, participants in different COST projects have fewer common points of contact than participants in different FP6 projects.

It should be stressed that there is disagreement in the literature as to whether small world networks really do encourage innovation, critical testing of ideas and/or communication. There are valid arguments and evidence pointing in both directions. But it is striking that the FP6 networks differ from other types of network in the extent to which small worlds characteristics appear in the data. However, the greater degree of clustering, the richness of connections and the short 'path lengths' among participating institutions should not be interpreted as directly transforming the entire ERA. Any public research programme reaches only a fraction of the research community, and FP6 reaches a smaller fraction than its predecessors. It may be that some of the increase in density and richness of connections reflects this restriction to a group of organisations that may have already had close connections. However, this alone cannot explain the bulk of the findings.

A final observation is that the world of FP research constitutes a small world – in the informal sense – of its own. The strategic objectives pose clear questions requiring novel combinations of research approaches and organisations. Some of the Instruments reinforce this by providing strong incentives for active collaboration.

## 3.4  Participants expect significant value from participating in FP6

The initial findings using data from surveys conducted for the European Commission by UK-based Technopolis and further validated by the 2004 IST Impact Study for Microelectronics & Microsystems, Health, and Mobile Communications.  These surverys show that the Framework programmes are attractive to people who expect to gain improved tools, methods or techniques through participation. Commercial products are not a major goal of participants: They are more attracted to the potential for improving their own internal operations and staying abreast of the latest research.  Findings also show that a large majority of participants expect to play a primary role in transferring the results of the research.

Value has traditionally been defined as "economic" and has been measured as such. However, within a knowledge-based society, the rising importance of intangible goods such as trust and other forms of social capital has expanded the definition of value to include both tangible and intangible value.   Intangible exchanges, such as knowledge and informational exchanges, are informal and difficult to measure – yet they are critical to creating value. Knowledge networks contribute to collaboration and build relationships. Intangible value is embedded in the expert knowledge of the FP6 participants.

The intangible value that is generated as outputs takes two forms:

- Knowledge itself expressed as new ideas and methods that are published, diffused, and influential throughout the research community.

- Social cohesion, which is expressed in the fabric of genuinely collaborative networks where different institutions are working together, as opposed to simply fulfilling their part of a contract.

It appears from the data that knowledge acquisition is a significant motive for people to participate in FP6 projects. The indicators from respondents suggested that Framework Programmes are a significant source of new knowledge for participants. Further the ratio of new to continuing partnerships points to a healthy turnover of knowledge: Not so ingrown as to stifle innovation yet not so open as to create chaos or lose continuity from one FP to the next. The indicator that shows 75 percent of the respondents feel they play a major or primary role in transferring the results of the research is a very positive indicator for knowledge diffusion. This suggests that knowledge is not just flowing in to the project but is also flowing outward in a substantial way.

Obtaining knowledge leadership through new tools and processes is a very significant motive for those participating in the FP6 projects. With 67 percent of participants expecting to gain improved tools, methods or techniques the innovations of most interest are those that are internally focused on how people approach their work rather than being externally focused on generating new products and services. Although not a major motivator, many participants are expecting to see new commercial products and improved scientific or industrial processes as an outcome.

This more internal focus on innovation suggests that evaluation approaches that seek only to identify commercial outcomes will not be able to discern the contribution of the programmes to intangible asset development. The core intangible asset categories are internal structures and systems, human competence and capability and the business relationships. These intangible outcomes are perhaps where the real value lies.

 # Lessons Learned: Insights, Opportunities, Challenges

## 4.1 Opportunities of Networked Research are Significant

The Sixth Framework Programme designers made a conscious effort to push for a more focused approach on denser interconnections. This meant larger projects (since within-project connections are stronger than among-project ones) and fewer participants (to ensure a 'small world' (in the lay sense) feeling. Other objectives were to include a more diverse sample of organisations and disciplines and to create new semantic clusters. The new Instruments and Strategic Objectives embody this; the FP put more direction on the formation and constitution of such groups, and less on their internal management. This replaced the old model of free entry and self-organisation, which led to a less dense, more fragmented structure seen throughout the rest of the ERA. On these terms, FP6 succeeded well - there are certainly fewer organisations, fewer projects per organisation and a denser set of interconnections. There were other effects as well. Some of these effects were hard-wired (larger projects, fewer of them) others were systemic or emergent consequences of the way the ERA responded (in making proposals and forming consortia) and the way the EC responded in selecting among the proposals. It is too soon to tell whether this more focused approach will have spillover effects to the rest of the ERA, whether the developments are good or bad on balance and how the effects divide across the stakeholder groups. But the importance of network structure as a diagnostic tool is established and any subsequent policy (whether to continue the FP6 intervention or not) must track these changes and develop links between network structures and more traditional (output/outcome-orientated) evaluation measures.

Network analysis provides a unique set of insights into the dynamics of research collaborations. The FP6 IST networks evaluated in this study appear to be providing an integrating function by drawing together otherwise fragmented players and isolated knowledge holders into a European-wide system. This is one of the goals of the IST RTD programme, and the network structures suggest that this goal is being reached. In addition, the analysis undertaken for the case studies suggests that the European-wide networks are more able to exploit global information for use within Europe. The FP6 networks are attracting knowledge leaders to work within the FP6 projects: the broader networks enabled by the redesigned Instruments makes this information widely available through the network to more participants than has been the case in other Frameworks.

The participants expect to develop new tools and gain access to world-class knowledge as a result of being involved in FP6.

However, the network effects that enable the FP6 programme to create greater connections among participants could be achieved with smaller projects than those created within FP6. The network benefits of interconnectivity are a feature of FP6 that is different from earlier Frameworks and that may be a positive influence in knowledge exchange. The very large projects may be creating an excess of connections that could actually work against the desired effect. The small world combination of active communication (short path lengths) and mutual exchange (high clustering) could be achieved with a set of overlapping small projects. This would have the added advantage of increasing the inclusion and standing of smaller participants and new entrants and could be realised by including a 'portfolio' or balance component to the individual screening when selecting proposals for funding.

The network approach to evaluation provides information on the dynamics of exchange within FP6 collaborations that would not be available using other kinds of evaluation tools. It provides a view of the ability to interconnect, to access state-of-the-art knowledge, and to connect with complementary research institutions. These insights could not be gained based on anecdotal reporting of individual Framework participants. In addition, the network approach to the role of large organisations, and their role in structuring collaboration for smaller organisations is also not an insight that could be readily gained from interviewing participants.

The Nautillus software tool also offers new ways of managing research. Using the networking software it is possible to see the impact of funding decisions upon links among proposed research partners. This would facilitate planning that carefully considered the number and extent of linkages among different participants. It could help balance networks to achieve diverse goals such as geographic spread and SME participation without losing the benefits of self-organisation that adds strength to the Framework programme.

## 4.2    Challenges of Managing Networks are also Significant

From the methodological point of view, a number of lessons emerge from this study. Research collaboration was viewed as being best achieved by facilitating self-organising networks, and networking became a means and an end of the FP6 programme. As the enthusiasm for networked structures has grown (not just in the EC, but in many venues), there has been a growing literature on network structure, the conduct of participants and the resulting performance of societal institutions. Part of this literature derived from considerations and studies of physical networks or networks that form through systemic interactions operating on random encounters. Clearly, these are not the same as the networks that grow based on EC incentives or other forms of rational or conscious choice. Still, the literature is not developed enough to differentiate among these different types of networks, so more research is needed to gain fuller insight when applying concepts and measurements derived from physical networks to apply them to observed networks where membership is based on choice.

To extend the utility of existing evaluation tools, ongoing monitoring studies should track network structure changes and correlate them to other measures – especially output-orientated measures. Also, a careful study should be made of the harder-to-observe network layers (e.g. personnel mobility, networks among researchers rather than institutions), the internal structure of the projects (in other words, not to treat them as stars or clusters, and also to take account of link strength and direction).

In terms of the structure of the Framework Programme, it seems likely that specific Instruments and funding arrangements produced selection effects (in other words, they influence who participates from the wider ERA) that can be seen in the characteristics of the participants (by size and public/private). They are also likely to have produced incentive effects (in other words why and how they participate) that have implications for both the structure and performance of internal (to projects) networks - but these cannot be observed from available data. The policy recommendation is to take this seriously: members of the ERA differ in the extent to which they rely on the FP as a source of business. It is possible to interpret the data as showing that FP6 may have encouraged the formation of an even tighter 'inside group' than previous FPs. This is not necessarily bad - others will then preferentially link to this core, and the good structural effects may spread to the rest of the ERA in that manner. But it should certainly be validated by further (tracking) study and reflected in policy. If policy makers wish to focus ERA efforts, design should take account of the 'cusp' presented by co-funding and large projects. Firms with a choice will split between those who respond as expected (e.g. by generating complementary outside research and links) and those who simply abandon the enterprise.

# Selected References